\documentclass[conference]{IEEEtran}
\IEEEoverridecommandlockouts
\usepackage{comment}
\usepackage{cite}
\usepackage{amsmath,amssymb,amsfonts}
\usepackage{algorithmic}
\usepackage{graphicx}
\usepackage{textcomp}
\usepackage{xcolor}
\def\BibTeX{{\rm B\kern-.05em{\sc i\kern-.025em b}\kern-.08em
    T\kern-.1667em\lower.7ex\hbox{E}\kern-.125emX}}
\begin{document}

\title{An Inception-Residual-based Architecture with Multi-objective Loss for Detecting Respiratory Anomalies
}
\author{Dat~Ngo$^{1}$, 
             Lam~Pham$^{2}$, 
             Huy~Phan*$^{3}$,
             Minh~Tran$^{4}$,
             Delaram~Jarchi$^{1}$,
             \c{S}efki~Kolozali$^{1}$
             \\ \\
1. School of Computer Science and Electronic Engineering, University of Essex, UK.\\
\{dn22678, delaram.jarchi, sefki.kolozali\}@essex.ac.uk \\            
2. Center for Digital Safety \& Security, Austrian Institute of Technology, Austria. \{lam.pham@ait.ac.at\} \\

3. Amazon Alexa, Cambridge, MA, US. \{huypq@amazon.co.uk\}  \\

4. Nuffield Department of Clinical Neurosciences, University of Oxford, UK. \{minh.tran@ndcn.ox.ac.uk\}\\

\thanks{(*) The work was done when Huy Phan was at the School of Electronic Engineering and Computer Science, Queen Mary University of London, UK, and prior to joining Amazon Alexa.}.

}

\maketitle

\begin{abstract}
This paper presents a deep learning system applied for detecting anomalies from respiratory sound recordings. Initially, our system begins with audio feature extraction using Gammatone and Continuous Wavelet transformation. This step aims to transform the respiratory sound input into a two-dimensional spectrogram where both spectral and temporal features are presented. Then, our proposed system integrates Inception-residual-based backbone models combined with multi-head attention and multi-objective loss to classify respiratory anomalies. 
Instead of applying a simple concatenation approach by combining results from various spectrograms, we propose a Linear combination, which has the ability to regulate equally the contribution of each individual spectrogram throughout the training process. To evaluate the performance, we conducted experiments over the benchmark dataset of SPRSound (The Open-Source SJTU Paediatric Respiratory Sound) proposed by the IEEE BioCAS 2022 challenge. As regards the Score computed by an average between the average score and harmonic score, our proposed system gained significant improvements of 9.7\%, 15.8\%, 17.8\%, and 16.1\% in Task 1-1, Task 1-2, Task 2-1, and Task 2-2, respectively, compared to the challenge baseline system. Notably, we achieved the Top-1 performance in Task 2-1 and Task 2-2 with the highest Score of 74.5\% and 53.9\%, respectively.
\end{abstract}


\section{Introduction}
\label{intro}

According to statistics from the World Health Organization (WHO)~\cite{who_1}, global mortality caused by respiratory diseases including tuberculosis, asthma, chronic obstructive pulmonary disease (COPD), and lower respiratory tract infection (LRTI) has reached an alarming number of 6.2 million at the time of writing. Furthermore, the heavy workload on the healthcare systems is currently not uncommon in many countries as the number of patients demanding on examining their respiratory sound patterns is increasing while the number of clinicians is limited. At the early stage, respiratory diseases cause destruction or blocking of the airways of the lung, which lead to the limitation of airflow during the inhalation and exhalation phases. Therefore, respiratory sound analysis by machine learning or deep learning methods has recently attracted much attention. They can be used to automate the processing of longitudinal data recorded outside the clinical environment and assist clinicians to make an improved and informed decision to detect precisely different respiratory sound patterns in a scalable, noninvasive, and time-saving manner. Regarding machine learning methods~\cite{lung_hmm_02,lung_svm_01, lung_tree_18}, respiratory sounds are first transformed into Mel-frequency cepstral coefficient (MFCC), referred to as hand-crafted features. Then, conventional machine learning models such as Hidden Markov Model~\cite{lung_hmm_02}, Support Vector Machine~\cite{lung_svm_01}, and Decision Tree~\cite{lung_tree_18} explore these handcrafted features to classify respiratory anomalies. Meanwhile, respiratory sounds are transformed into a two-dimensional spectrogram, where both temporal and spectral information are fully represented in a wider time context in deep learning-based systems. These spectrograms are then fed into powerful network architectures such as convolutional neural network (CNN) based architectures~\cite{pham2021cnn, ngo2021deep} or recurrent neural network (RNN) based architectures~\cite{ic_cnn_19_iccas, perna2019deep} for classification. 
Compared to the performances between machine learning and deep learning-based approaches, the latter demonstrates more effectiveness for detecting respiratory anomalies~\cite{pham2021cnn, 9871440, ngo2021deep}.

To leverage the deep learning-based approach, we propose in this paper a deep learning-based system for detecting anomalies in respiratory sounds. To demonstrate the performance, we evaluate our proposed system on the benchmark dataset of 2022 SPRSound~\cite{zhang2022sprsound}. Our contributions are as follows: (1) We investigated multiple spectrogram inputs extracted from Gammatone filter and Continuous Wavelet transformation, and then proposed an effective Linear combination method, which gathers complement information extracted from these individual spectrograms to form a comprehensively combined feature; (2) We also integrated multiple techniques of Inception-residual-based network architecture, multi-head attention, and multi-objective loss into our proposed system to fully exploit the combined feature and gained competitive results compared to state-of-the-art systems.

The rest of this paper is organized as follows. Section~\ref{related_work} presents related work and our motivation. Section~\ref{dataset} describes the dataset that is used to evaluate our proposed method. Section~\ref{proposedframework} details our proposed system. Experiments and results are evaluated in Section~\ref{experiment}. The conclusion is given in Section~\ref{conclusion}.

\section{Related work and Motivation}
\label{related_work}
As mentioned in Section~\ref{intro}, we now analyze the related work of two main factors in a deep learning-based system for detecting anomalies from a respiratory sound: The front-end feature extraction using spectrograms and the back-end deep neural network architecture. Next, we then propose the main techniques which are applied to our proposed system to improve the performance.

Regarding the feature extraction, almost all of the state-of-the-art systems for detecting anomalies from respiratory sounds use Mel-filter~\cite{li2022improving, nguyen2022lung}  or short-time
Fourier transformation (STFT)~\cite{chen2022classify, jung2021efficiently} to generate spectograms. However, as using the fixed window size, these types of spectrograms might not be able to well present discrimination between normal and abnormal respiratory sounds. To overcome this issue, we propose an alternative way of using multiple spectrograms with Wavelet-based spectrograms~\cite{pham2021inception} and Gammatone-filter-based spectrogram~\cite{pham2021cnn, 9871440, ngo2021deep} for this work. While the former generates a better multi-resolution analysis thanks to its suitability in adjusting both temporal window length and the wide frequency range across the length, the latter shows great ability to capture the energy of the sound signals in a lower frequency band, which is strongly related to the nature of respiratory sounds. 

As regards the deep neural network architecture, our prior work in~\cite{pham2021inception} using an Inception-based network and ensemble of models with multiple spectrograms~\cite{pham2022ensemble, pham2021inception, jung2021efficiently} have witnessed the improvement to detect respiratory anomalies. Therefore, these inspire us to propose a new architecture of Inception-residual-based backbone model supported by multi-head attention and multi-objective loss to train multiple spectrograms at the same time. Additionally, rather than applying a naive ensemble of results from different spectrograms, a Linear combination, a part of our proposed model, is proposed to regulate equally the contribution of each individual spectrogram and generate a combined feature representing aspects from all spectrogram inputs during the training course.

\section{Dataset and tasks defined}
\label{dataset}
\subsection{SPRSound dataset}

In this work, we evaluate our systems on the benchmark 2022 SPRSound: Open-Source SJTU Paediatric Respiratory Sound database, which was collected in the Shanghai Children’s Medical Center (SCMC), China~\cite{zhang2022sprsound}. 
The dataset consists of 2683 audio recordings collected from 292 patients (whose age is from 1 month to 18 years old). 
After being carefully inspected by experts, the quality of each recording was labeled as \textit{Poor Quality (PQ)}, \textit{Normal (N)}, \textit{Continuous Adventitious Sound (CAS)}, \textit{Discontinuous Adventitious Sound (DAS)}, and \textit{CAS and DAS (CD)} with the recording numbers of 187, 1785, 233, 347, and 131, respectively.
For each audio event in a recording, the onset (i.e. starting time) and offset (i.e. ending time) are professionally labeled by respiratory experts.
There are total of 6887 audio events of \textit{Normal (N)}, 53 audio events of \textit{Rhonchi (Rho)}, 865 audio events of \textit{Wheeze (W)}, 17 audio events of \textit{Stridor (Str)}, 66 audio events of \textit{Coarse Crackle (CC)}, 1167 audio events of \textit{Fine Crackle (FC)} and 34 audio events of \textit{Wheeze and Crackle (B)}. 

Overall, there is an issue of imbalance between audio events as well as entire recordings.
Additionally, these audio recordings and events show various duration ranging from 0.304\,s-15.36\,s and 0.126\,s to 7.152\,s, respectively, which makes the 2022 SPRSound dataset more challenging.

\subsection{Tasks Definition}

Given the SPRSound dataset, the IEEE BioCAS 2022 challenge proposed two main tasks. 
First, \textbf{Task 1}, which aims to focus on the sound event level, is divided into 2 sub-tasks: Task 1-1 and Task 1-2. 
While Task 1-1 is to classify the respiratory sound events as Normal and Adventitious, Task 1-2 aims to classify these events into \textit{N}, \textit{Rho}, \textit{W}, \textit{Str}, \textit{CC}, \textit{FC}, or \textit{B}.
Second, \textbf{Task 2} focuses on the entire recording, which is also separated into Task 2-1 and Task 2-2. 
In particular, Task 2-1 accounts for classifying the respiratory recordings as Normal, Adventitious, and Poor Quality. 
Meanwhile, Task 2-2 is a multi-class classification, where the respiratory recordings are classified into \textit{N}, \textit{CAS}, \textit{DAS}, \textit{CD}, or \textit{PQ}. 
To adhere to the evaluation metrics in the IEEE BioCAS 2022 challenge, every task and its sub-tasks in this paper are evaluated by \textit{sensitivity} (SE), \textit{specificity} (SP), \textit{average score} (AS), \textit{harmonic score} (HS), and the average of AS and HS (Score) as follows:

\begin{equation}
SE = \frac{ \Sigma T_i}{\Sigma N_i}
 \end{equation}

\begin{equation}
SP = \frac{T_N}{N_N},
 \end{equation}

where $T_i$ and $T_N$ denote the number of correctly classified
samples of target class $i$, and class Normal, respectively. $N_i$ and $N_N$ are the numbers of all samples of target class $i$ and class Normal, respectively.

Next, the average score (AS), the harmonic score (HS), and the Score are computed by:

\begin{equation}
AS =  \frac{SE+SP}{2}
 \end{equation}

\begin{equation}
HS =  \frac{2.SE.SP}{SE+SP}
 \end{equation}

\begin{equation}
Score =  \frac{AS+HS}{2}
 \end{equation}

\section{The Proposed System}
\label{proposedframework}

\begin{figure}[t]
	\centering
	\centerline{\includegraphics[width=0.9\linewidth]{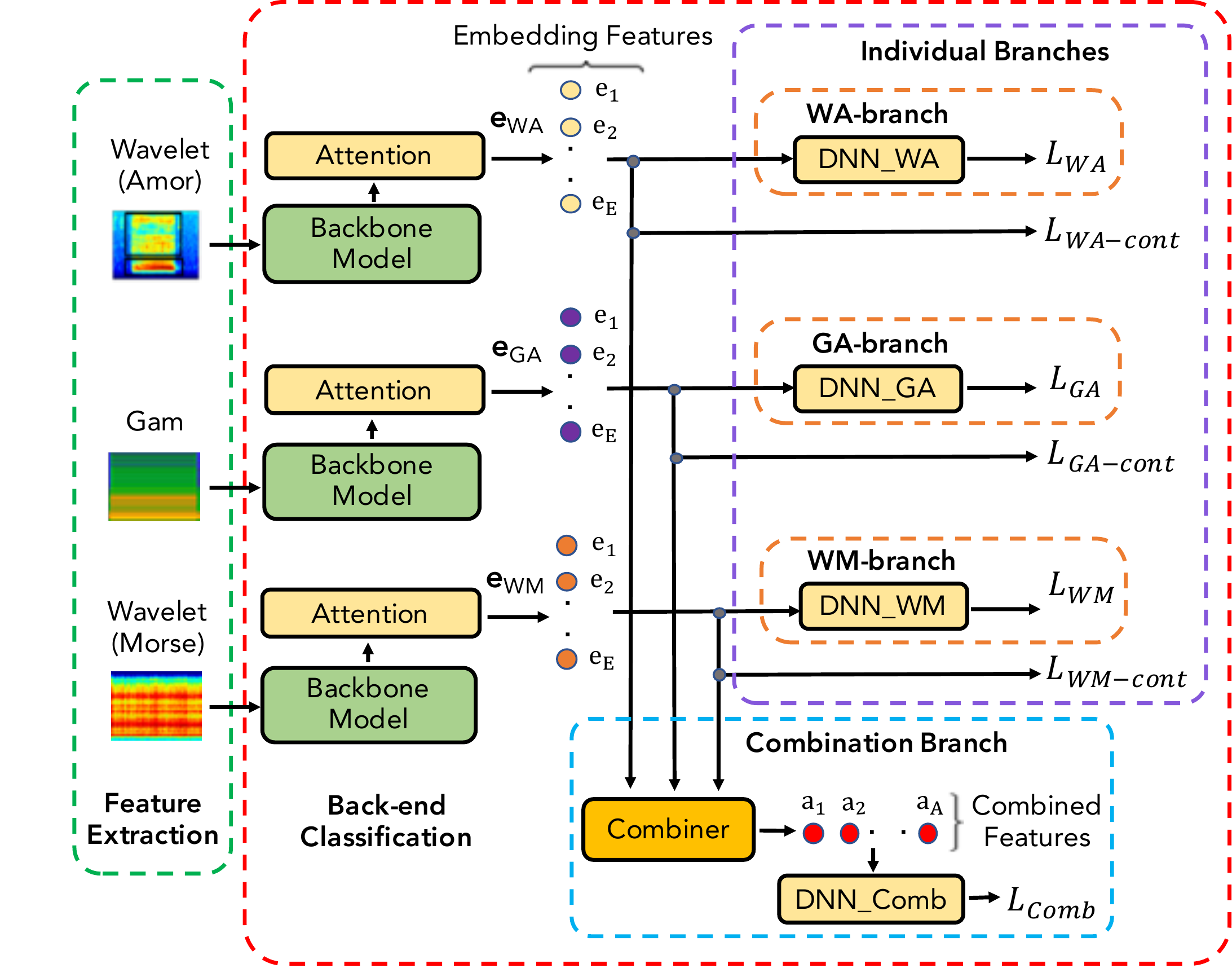}}
	 \vspace{0 cm}
	\caption{The high-level architecture of our proposed system for detecting anomalies in respiratory audio recording.}
	\vspace{-0.4 cm}
	\label{fig:overall_framework}
\end{figure}	

Overall, our proposed system for detecting anomalies in respiratory sounds as shown in Fig.~\ref{fig:overall_framework} consists of two main steps: low-level spectrogram feature extraction and back-end classification.

\begin{figure*}[th]
	\centering
	\centerline{\includegraphics[width=1\linewidth]{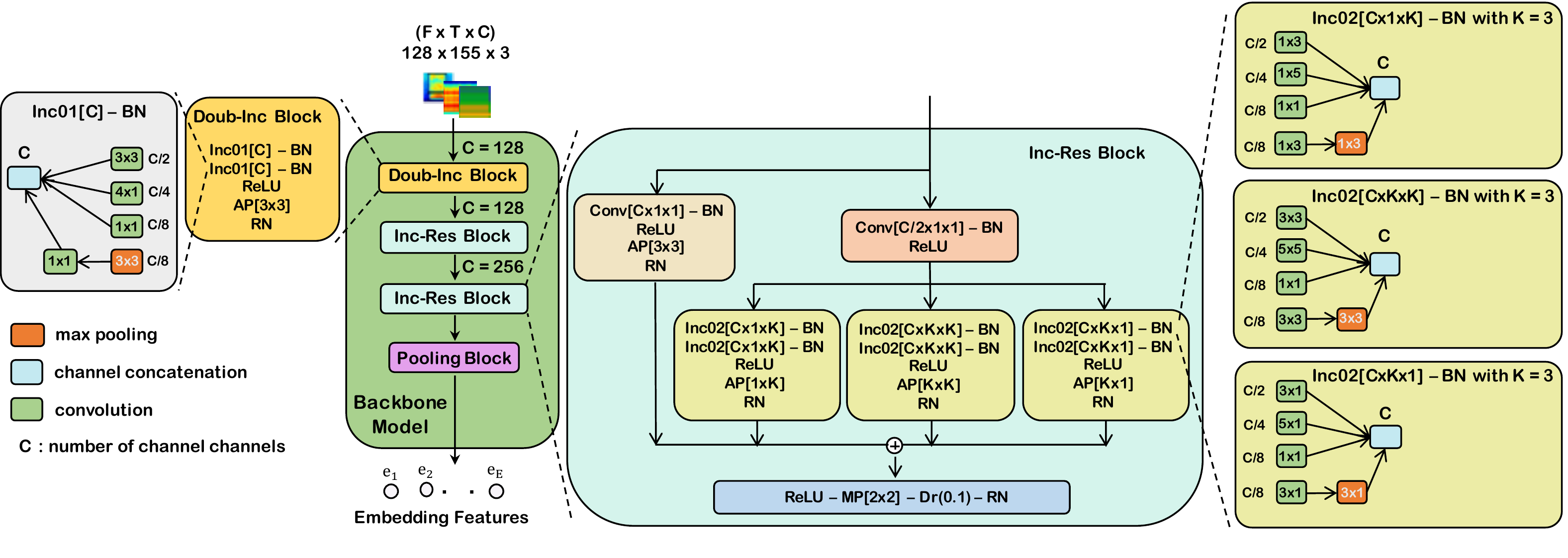}}
	 \vspace{-0.3 cm}
	\caption{The detailed architecture of the Backbone Model used in the proposed system}
	\vspace{-0.4 cm}
	\label{fig:backbone_archi}
\end{figure*}

\subsection{The low-level spectrogram feature extraction}

At the sound event level, the respiratory events evaluated in Task 1 are re-sampled to 4 kHz as all abnormal sounds have their frequency bands located around 60-2000 Hz~\cite{zhang2022sprsound}. 
Next, re-sampled respiratory events presenting different recording lengths are then duplicated to obtain the same length of 10 seconds. 
A band-pass filter of 60-2000 Hz is then applied to suppress background noise in every respiratory event. 
After that, these respiratory events are transformed into a two-dimensional spectrogram by using Gammatone~\cite{aud_tool} and Continuous Wavelet transformations. For the Continuous Wavelet transformations, we use Amor and Morse as the Wavelet mother functions. 
As a result, we obtain three spectrograms from one sound event.
Finally, all spectrograms are scaled into the size of $128{\times}155$ (i.e. frequency bands${\times}$the
number of time frames). 

At the recording level for evaluation in Task 2, the entire audio recordings are duplicated to make sure that all audio recordings have a consistent duration of 15.36 seconds in accord with the longest recording. Similar to Task 1, these audio recordings are then transformed into Gammatone spectrogram and Wavelet-based spectrograms. 
Finally, entire spectrograms are scaled to the sizes of $128{\times}512$ (i.e. the longer duration of the entire recording results in more time frames compared to the event level). 

After generating spectrograms, we apply three data augmentation methods on the spectrograms on both event and entire recording levels to deal with the imbalanced data issue mentioned in Section~\ref{dataset}. 
First, we randomly oversample spectrograms to make sure that the number of spectrograms per class is equal in each batch size. Second, the mixup data augmentation \cite{tokozume2017learning} is applied to increase the variation of the training data and to enlarge Fisher’s criterion (i.e. the ratio of the between-class distance to the within-class variance in the feature space). Third, augmented spectrograms in every batch are randomly cropped with a reduction of 10 bins on both time and frequency dimensions to motivate learning toward the partial losses of information at each dimension~\cite{park2019specaugment}.

\subsection{The back-end classification}
The back-end classifier comprises four main branches. 
The first three branches comprise \textbf{WA-branch}, \textbf{GA-branch}, and \textbf{WM-branch}, which aim to focus on exploring individual spectrogram inputs as shown in the upper part of Fig.~\ref{fig:overall_framework}. We denote the spectrogram generated by Gammatone transformation as GA, whereas we mark WA and WM as obtained spectrograms from Amor and Morse on the Wavelet mother functions in Continuous Wavelet Transformations, respectively. Meanwhile, the final branch, referred to as \textbf{Combination branch}, collects embedding features extracted from the first three individual branches (i.e. the outputs of Attention blocks) to derive a Combined feature as shown in the lower part of Fig.~\ref{fig:overall_framework}. 

Individual branches of \textbf{WA-branch}, \textbf{GA-branch}, and \textbf{WM-branch} comprises a Backbone Model followed by an Attention block and a DNN-based block (i.e. DNN\_WA, DNN\_GA, or DNN\_WM). In particular, the detailed architecture of the Backbone Model is illustrated in Fig.~\ref{fig:backbone_archi}. 
The proposed Backbone Model based on Inception-residual-based network architecture comprises four blocks: one Doub-Inc Block, two Inc-Res Blocks, and one Pooling Block. 
The Doub-Inc Block performs two Inc01 layers followed by Batch Normalization (BN), Rectified Linear Unit (ReLU), Average Pooling (AP[Kernel Size]), Dropout (Dr(Drop Ratio), and Residual Normalization (RN( $\lambda$ = 0.4)) inspired from\cite{kim2022qti}).  
The Inc01 layer is a variant of the naive inception layer~\cite{szegedy2015going} with fixed kernel sizes of [3×3], [1×1], and [4×1].
Regarding two Inc-Res Blocks, they share the same architecture, but channel numbers increase from 128 to 256 to form a deeper view of the channel dimension as inspired by our prior work in~\cite{pham2021inception}. 
The detailed structure for every Inc-Res Block is described in the middle part of Fig.~\ref{fig:backbone_archi}, which features layers of Inc02[Channel×Kernel Size], Conv, BN, Dr, ReLU, AP, Max Pooling (MP[Kernel Size]), and RN( $\lambda$ = 0.4).
Notably, three different Inc02[Channel×Kernel Size] layers, which present different kernel sizes defined as [K×1], [K×K], and [1×K], are used in Inc\_Res Block. 
The value of K is changed as details on the right-hand side of Fig.~\ref{fig:backbone_archi}. 
Following each Inc02 layer, an AP layer is applied before adding sub-branch results together. Furthermore, we propose to apply RN to form the Inc-Res Block as RN can be used as instance normalization with a shortcut path to eliminate the influence of unnecessary responses from some frequency bands during training (i.e. these influences caused by noises from environments like coughing, motion artifacts, and intestinal and physiological sounds when respiratory sounds are recorded). This idea strengthens the backbone to learn effectively not only the widespread frequency bands but also the distribution of energy in certain frequency bands from different types of anomalies in lung sound. The Pooling Block makes use of global pooling layers to extract three features from the second Inc-Res Block: (1) global average pooling across the channel dimension, (2) global max pooling across the temporal dimension, and (3) global average pooling across frequency dimensions.

As the work in~\cite{vaswani2017attention} indicates that each attention head learns a different set of weight matrices and when joining many self-attention heads together, it forms a multi-head self-attention layer which helps to learn an input feature map better. Therefore, in this work, we propose that the output of the Backbone Model (i.e. The output of the Pooling Block in Fig.~\ref{fig:backbone_archi}) is then presented to the Attention block as shown in Fig.~\ref{fig:overall_framework}. 
At each Attention block, we apply three multi-head attention layers~\cite{vaswani2017attention} on three dimensions of frequency, time, and channel to fully exploit the embeddings extracted from different aspects of the spectrograms (i.e. specific embeddings extracted from different spectrograms). Each multi-head attention layer is configured to have 16 as the number of heads and 32 as the key dimension.
The output of each multi-head attention layer is a one-dimensional embedding feature (i.e. the embedding feature is described as a vector $\mathbf{e}[e_{1}, e_{2}, ..., e_{E}]$ in Fig.~\ref{fig:overall_framework}).

The embedding features from Attention Blocks are finally fed into individual DNN\_WA, DNN\_GA, or DNN\_WM blocks for classification. 
The blocks of DNN\_WA, DNN\_GA, and DNN\_WM share the same network architecture which comprises two dense layers. 
The first dense layer comprises a fully connected layer (FC[C = $E$]) layer followed by BN, ReLU, and Dr, where $E$ is the dimensional length of the embedding features.
Meanwhile, the second dense layer comprises a fully connected layer (FC[C = $T$]) followed by a Softmax, where $T$ is defined according to the number of target classes.

\textbf{Combination branch architecture:} This final branch as shown in the lower part of Fig.\ref{fig:overall_framework} comprises two main blocks: Combiner and DNN\_Comb.
The Combiner block gathers embedding features $\mathbf{e}[e_{1}, e_{2}, ..., e_{E}]$ outputted from Attention blocks to obtain a Combined feature $\mathbf{a}[a_{1}, a_{2}, ..., a_{E}]$.
As the use of multiple spectrograms has shown improvement as mentioned in Section~\ref{related_work}, we propose this Combiner block. We assume that each spectrogram contains distinct features, and then the Combined feature from the Combiner block are potentially better representing features from all the spectrograms for detecting anomalies in respiratory sounds. In this work, we propose two combination methods. The first method is simple concatenation. 
In the second method, we assume that the feature embeddings $\mathbf{e}[e_{1}, e_{2}, ..., e_{E}]$  from each spectrogram have a linear relationship across their dimensions. 
We then derive a data-driven combination method called Linear combination. 
This Linear combination is proposed to combine the three high-level features with trainable weight and bias:

\begin{equation}
\label{eq:1}
\begin{aligned}
\textbf{a}_{Linear-combined} = ReLU(\textbf{e}_{WA}\textbf{w}_{WA} + \\
\textbf{e}_{GA}\textbf{w}_{GA} + \textbf{e}_{WM}\textbf{w}_{WM} + \textbf{w}_{bias}),
\end{aligned}
\end{equation}

where ${\textbf{w}}_{WA/GA/WM/bias}$[$w_{1}$, $w_{2}$, ..., $w_{E}$] are the trained parameters. The Combined feature $\textbf{a}$  are finally fed into the DNN\_Comb block for classification. The DNN\_Comb block presents the same architecture as DNN\_GA, DNN\_WA, or DNN\_WM blocks.
\begin{table*}[th]
    \caption{Performance comparison on different systems on the official test set in SPRSound dataset 2022} 
        	\vspace{-0.2cm}
    \centering
    \scalebox{0.95}{
    \begin{tabular}{|c| c c|c c|c c| c c|c c|c c|} 
        \hline 
             
             \textbf{System}&\multicolumn{2}{|c|}{\textbf{WA-branch}} &\multicolumn{2}{|c|}{\textbf{GA-branch}} &\multicolumn{2}{|c|}{\textbf{WM-branch}} & \multicolumn{2}{|c|}{\textbf{System I}} &\multicolumn{2}{|c|}{\textbf{System II}} &\multicolumn{2}{|c|}{\textbf{System III}} \\
    
        \hline 
             &SE/SP  &AS/HS  &SE/SP  &AS/HS   &SE/SP  &AS/HS &SE/SP  &AS/HS  &SE/SP  &AS/HS   &SE/SP  &AS/HS  \\
             \hline
             Task 1-1 &77.3/80.7  &78.9/78.9  &87.9/72.1 &80.0/79.2  &74.5/88.6  &81.5/80.9  &70.4/87.9  &81.2/80.6  &79.4/85.9 &82.7/82.5  &\textbf{84.4/85.5}  &\textbf{84.9/84.9}\\
             Task 1-2 &49.4/87.1  &68.2/63.0  &55.2/84.8 &70.0/66.9  &66.4/70.3  &68.3/68.2  &56.1/89.5  &72.8/69.0  &65.3/87.7 & 76.5/74.9  &\textbf{67.8/88.3}  &\textbf{78.1/76.7}\\
             Task 2-1 &46.5/79.1  &62.8/58.6  &70.3/50.6 &60.5/58.8  &62.4/64.4  &63.4/63.4  &58.1/72.2  &65.2/64.4  &65.6/77.1 &71.4/70.9  &\textbf{70.4/78.9}  &\textbf{74.7/74.4}\\
             Task 2-2 &22.6/71.1  &46.9/34.3  &24.8/72.1 &48.5/36.9  &26.2/74.2  &50.2/38.7  &19.1/91.7  &55.4/31.6  &25.2/83.6 &54.4/38.7  &\textbf{36.1/80.1}  &\textbf{58.1/49.8}\\
            
       \hline 
    \end{tabular}
                       }
        \vspace{-0.3cm}
    \label{table:framework_comp} 
\end{table*}
\textbf{Training loss functions:} we propose different loss functions to train the proposed system. 
In particular, we use two loss functions of KL-loss and contrastive loss~\cite{chopra2005learning} for each Spec-Ind branch to learn individual spectrogram input. First, as using mixup data augmentation, the labels are not one-hot format.
Therefore, we use Kullback-Leibler (KL) divergence loss in the proposed networks as shown in Eq. (\ref{eq:kl_loss}) below:

\begin{align}
\label{eq:kl_loss}
Loss_{KL}(\Theta) = \sum_{n=1}^{N}\mathbf{y}_{n}\log(\frac{\mathbf{y}_{n}}{\mathbf{\hat{y}}_{n}})  +  \frac{\lambda}{2}||\Theta||_{2}^{2},
\end{align}

where \(Loss_{KL}(\Theta)\) is KL-loss function, $\Theta$ describes the trainable parameters of the network, $\lambda$ denotes the $\ell_2$-norm regularization coefficient experimentally set to 0.0001, \(N\) is the batch size, $\mathbf{y_{n}}$ and $\mathbf{\hat{y}_{n}}$  are the ground truth and the network output, respectively. We set the learning rate to 0.0001 and Adam method \cite{Adam} is  applied for optimization. While KL-loss functions of $L_{WA}$, $L_{GA}$, $L_{WM}$ help to classify target classes in each individual spectrogram branch, KL-loss of $L_{Comb}$ is used in the Combination branch.

Second, we also apply contrastive loss in our training process.
We manage to have a pair of spectrograms $(S_i, S_j)$ and a label $Y$. The label is equal to 1 if the spectrograms are in the same class and 0 otherwise. To extract embedding features of each spectrogram, we use the Backbone model $f$ that encodes the spectrogram $S_i$ and $S_j$ into an embedding space where $\mathbf{e_i} = f(S_i)$ and $\mathbf{e_j} = f(S_j)$. The contrastive loss is defined as:

\begin{align}
\label{eq:const_loss}
L_{Cont} = Y*d^{2} + (1-Y)max(margin - d,0)^{2},
\end{align}

where $d$ = $||\mathbf{e_i} - \mathbf{e_j}||_{2}$ is the Euclidean distance between embeddings and $margin$ is set to 1. In every spectrogram branch, the contrastive loss functions of $L_{WA-Cont}$, $L_{GA-Cont}$, and $L_{WM-Cont}$ are applied to embedding features, which helps to maximize the Euclidean distance between embeddings from different classes while minimizing the Euclidean distance between embeddings from the similar classes. 

Eventually, the multi-objective loss is computed as:

\begin{equation}
\begin{aligned}
L_{total} = \alpha (L_{WA} + L_{GA} + L_{WM}) +  \beta (L_{Comb}) + \\
\gamma(L_{WA-Cont} + L_{GA-Cont} + L_{WM-Cont} ),
\end{aligned}
 \end{equation}

where $\alpha$, $\beta$, and $\gamma$ are loss weight ratios that are used to manage the contribution of every single objective loss.

 \section{Experiments and results}
\label{experiment}



As we use multiple techniques of multiple branches, multi-head attention, features combinations, and multi-objective loss function to construct our proposed system as shown in Figure~\ref{fig:overall_framework}, we then evaluate the individual role of certain techniques by proposing four system variants:

\textbf{Individual branch:} This system only uses one backbone model and one of three branches: WA-branch, GA-branch, or WM-branch. This system is used to evaluate the contribution of every individual spectrogram.

\textbf{System I:} This system uses all four branches, the concatenation method is used in the Combiner block, but does not use multi-head attention and contrastive losses. The values of loss weight ratios of $\alpha$, $\beta$ and $\gamma$ is empirically set to 1/3, 1, and 0, respectively.

\textbf{System II:} This system reuses the settings of System I and Attention blocks are used. The values of loss weight ratios of $\alpha$, $\beta$ and $\gamma$ is empirically set to 1/3, 1, and 0, respectively.

\textbf{System III:} This system uses all techniques mentioned in this paper (i.e. Use Attention block, Linear combination method is used in Combiner block, both KL-loss and contrastive loss). The values of loss weight ratios of $\alpha$, $\beta$ and $\gamma$ is empirically set to 1/3, 1, and 1, respectively.

As Table~\ref{table:framework_comp} shows, in each individual spectrogram branch, tasks in the event level (i.e. Task 1-1, Task 1-2) has a better performance compared to tasks in the recording level (i.e. Task 2-1, Task 2-2). This can be explained as Task 1-1 and 1-2 only focus on certain events, while Task 2-1 and 2-2 focus on the entire recording including different types of events, and poor quality recordings are treated as one of the categories in Task 2-1 and Task 2-2. Among individual spectrograms, WM has the best performance with HS in Task 1-1, Task 1-2, Task 2-1, and Task 2-2 are 80.9\%, 68.2\%, 63.4\%, and 38.7\%, respectively. 

In Table~\ref{table:framework_comp}, System I with multiple spectrogram inputs indeed improves the performance in most tasks compared to that in individual WA-branch and GA-branch. However, Task 2-2 in System I witnesses an overfitting performance of 91.7\% in SP. Notablly, although System I uses multiple spectrograms with simple concatenation, there is even a downgrade in performance in Task 1-1 with lower AS/HS of 81.2\%/80.6\% compared to 81.5\%/80.9\% in the best single spectrogram WM-branch.

When the multi-head attention technique is applied as mentioned in System II, all performance criteria of SE/SP and AS/HS are improved compared with System I and single spectrogram branches. The issue of overfitting on SP in Task 2.2 of System I is also resolved. For instance, the performance of HS in Task 2-1 and Task 2-2 witnesses an improvement of 6.5\% and 7.1\%, respectively, compared to that in System I. 

\begin{figure}[t]
	\centering
		 \vspace{0.5 cm}
\centerline{\includegraphics[width=1\linewidth]{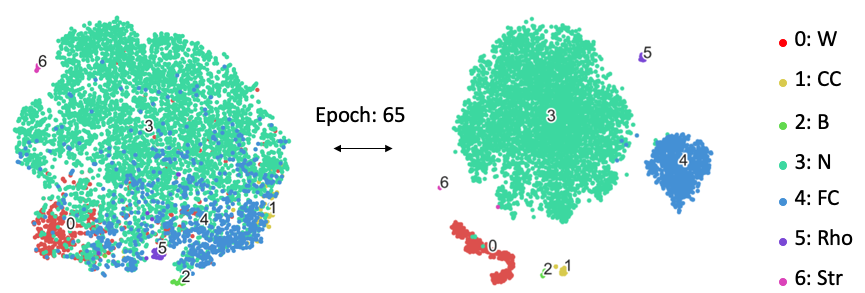}}
	\caption{The efficiency of applying contrastive loss in the training process on Task 1-2 (On the left: t-SNE map of embeddings extracted from 
 System III without contrastive loss. On the right: t-SNE map of embeddings extracted from System III with contrastive loss)}
	\vspace{-0.4 cm}
	\label{fig:contrastive_loss}
\end{figure}

Regarding System III, applying both Multi-head attention and a Linear combination of embedding features helps to further improve the performance and proves that System III achieves the best performance compared to the other evaluating systems. As shown in Table~\ref{table:framework_comp}, the highest AS/HS of 84.9\%/84.9\% are observed in Task 1-1. On Task 1-2, this system also surpasses other systems in terms of AS and HS, with the results being 78.1\% and 76.7\%. Notably, the superiority of this system is reflected as HS in Task 2-2 witnesses a significant improvement of 18.2\% and 11.1\% compared with that in System I and System II, respectively. This again demonstrates that multi-head attention and Linear combination play a significant role in increasing performance. In addition, the implementation of contrastive loss in System III helps to significantly reduce the training time and comprehensively discriminate the distribution of embedding features among target classes.
For instance, Fig.~\ref{fig:contrastive_loss} shows that 
System III with contrastive loss converges faster at the same epoch of 65 compared to the same system without using the contrastive loss in Task 1-2.

Compared to the Challenge baseline and top-3 systems submitted for the IEEE BioCAS 2022 challenge as shown in Table~\ref{table:backbone_results} (i.e. these systems just reported the Scores), we gain Scores of 84.9\% in Task 1-1 and 77.4\% in Task 1-2 while the highest Scores are 89.0\% and 82.0\%, respectively. Significantly, our proposed System III has demonstrated more efficiency than others as it outperforms the state-of-the-art in Task 2-1 and Task 2-2, achieving the highest Score of 74.5\% and 53.9\%. However, the low performances on Task 2-2 compared to Task 1-2 in Table~\ref{table:backbone_results} present a challenge for the task of multi-class classification on the entire recording level.

\begin{table}[t]
	\caption{Performance of our proposed system compared to others on the official test set in SPRSound dataset 2022 (Score(\%))}
        	\vspace{-0.2 cm}
    \centering
    \scalebox{0.95}{
    \begin{tabular}{|c| c c c c|} 
        \hline 
	   \textbf{Systems}    &\textbf{Task 1-1}  &\textbf{Task 1-2}   &\textbf{Task 2-1}  &\textbf{Task 2-2} \\
        \hline 
	    Challenge Baseline~\cite{zhang2022sprsound}          &75.2             &61.6            &56.7     &37.8  \\
        \hline
	    Top 1~\cite{li2022improving}                     &88.9             &\textbf{82.0}          &71.8     &\textbf{53.3}     \\
	    Top 2~\cite{zhang2022feature}                      &82.0             &74.3            &71.1     &53.1     \\
        Top 3~\cite{chen2022classify}                      &\textbf{89.0}             &80.0            &71.0     &36.0     \\
         \hline
	    Our System III       &84.9             &77.4           &\textbf{74.5}   &\textbf{53.9}  \\
          \hline
    
    \end{tabular}
    }
    \vspace{-0.4cm}
    \label{table:backbone_results} 
\end{table}

\section{Conclusion}
\label{conclusion}
We have presented a deep learning system, which integrated an Inception-residual-based network architecture with various techniques such as multiple spectrogram inputs, multi-head attention, multi-objective loss, and a Linear combination for detecting and classifying anomalies in respiratory sounds. The experiment results on the IEEE BioCAS 2022 challenge have indicated that multi-head attention and Linear combination play a significant role in increasing performance. While multi-head attention helped to extract good embedding from each individual spectrogram, the Linear combination combined and regulated effectively the contribution of each individual spectrogram. The efficacy has been demonstrated in our proposed system (i.e. System III) with a very competitive Score of 84.9\% in Task 1-1, 77.4\% in Task 1-2, and Top-1 performance in both Task 2-1 and Task 2-2 with 74.5\%, and 53.9\%, respectively.




\end{document}